\documentclass[pre,aps,twocolumn,amsmath,amssymb,superscriptaddress,floatfix]{revtex4-1}

\usepackage{graphicx}
\usepackage{amsmath}
\usepackage{xcolor}
\usepackage{todonotes}
\usepackage{menukeys}
\usepackage{hyperref}
\usepackage{acronym}
\usepackage{array,mathtools,amssymb,booktabs}
\usepackage{physics}
\AtBeginDocument{
\heavyrulewidth=.10em
\lightrulewidth=.05em
\cmidrulewidth=.03em
\belowrulesep=.65ex
\belowbottomsep=0pt
\aboverulesep=.4ex
\abovetopsep=0pt
\cmidrulesep=\doublerulesep%
\cmidrulekern=0.5em
\defaultaddspace=.5em
}

\usepackage{acronym}

\begin{document}
  \newlength\figurewide%
  \figurewide=.5\columnwidth%

\title{Neural network approach for the dynamics 
  on the normally hyperbolic invariant
  manifold of periodically driven systems}
\author{Martin Tsch\"ope}
\author{Matthias Feldmaier}
\author{J\"org Main}
\affiliation{%
Institut f\"ur Theoretische Physik 1,
Universit\"at Stuttgart,
70550 Stuttgart,
Germany}

\author{Rigoberto Hernandez}
\email[Correspondence to: ]{r.hernandez@jhu.edu}
\affiliation{%
Department of Chemistry,
Johns Hopkins University,
Baltimore, Maryland 21218, USA}
\date{\today}
\renewcommand{\vb}[1]{\boldsymbol{#1}}
\newcommand{\Ws}{\mathcal{W}_\mathrm{s}}
\newcommand{\Wu}{\mathcal{W}_\mathrm{u}}
\newcommand{\Wsu}{\mathcal{W}_\mathrm{s,u}}
\newcommand{\sno}[1]{_\mathrm{#1}}
\newcommand{\EQ}{Eq.}
\newcommand{\EQS}{Eqs.}
\newcommand{\FIG}{Fig.}
\newcommand{\FIGS}{Figs.}
\newcommand{\REF}{Ref.}
\newcommand{\REFS}{Refs.}
\newcommand{\SEC}{Sec.}
\newcommand{\SECS}{Secs.}
\newcommand{\eg}{e.\,g.}
\newcommand{\ie}{i.\,e.}
\newcommand{\cf}{cf.}

\begin{abstract}
  Chemical reactions in multidimensional systems are often described
  by a rank-1 saddle, whose stable and unstable manifolds intersect in
  the normally hyperbolic invariant manifold (NHIM).  Trajectories
  started on the NHIM in principle never leave this manifold when
  propagated forward or backward in time.
  However, the numerical
  investigation of the dynamics on the NHIM is difficult because of
  the instability of the motion.  We apply a neural network to
  describe time-dependent NHIMs and use this network to stabilize the
  motion on the NHIM for a periodically driven model system with two
  degrees of freedom. The method allows us to analyze the dynamics on
  the NHIM via Poincar\'e surfaces of section (PSOS) and to determine
  the transition state (TS) trajectory as a periodic orbit with the
  same periodicity as the driving saddle, viz.\ a fixed point of the
  PSOS surrounded by near-integrable tori.  Based on Transition State
  Theory and a Floquet analysis of a periodic
  TS trajectory we compute the
  rate constant of the reaction with significantly reduced numerical
  effort compared to the propagation of a large trajectory ensemble.
\end{abstract}
\maketitle

\preto\section\acresetall
\acrodef{NN}{neural network}
\acrodef{DS}{dividing surface}
\acrodef{TST}{Transition State Theory}
\acrodef{TS}{transition state}
\acrodef{LD}{Lagrangian descriptor}
\acrodef{NHIM}{normally hyperbolic invariant manifold}
\acrodef{TD}{time descriptor}
\acrodef{GP}{Gaussian Process}
\acrodef{PSOS}{Poincar\'e surface of section}

\section{Introduction}
\label{sec:Introduction}

In chemical reactions,
the precise separation between reactants and products is
a key to determining rate constants.
Usually, the boundary between these regions contains an
energetic barrier in phase space
---typically a rank-1 saddle---
to which an appropriate \ac{DS} can be 
{associated or} attached.
\ac{TST}~\cite{pitzer,%
  pechukas1981,%
  truh79,truh85,%
  hynes85b,berne88,nitzan88,%
  rmp90,truhlar91,truh96,truh2000,%
  Komatsuzaki2001,%
  pollak05a,%
  Waalkens2008,hern08d,Komatsuzaki2010,%
  hern10a,Henkelman2016}
uses the particle flux through a \ac{DS} to determine the rate of a
chemical reaction.

For an exact reaction rate, it is crucial to have a \emph{recrossing-free}
\ac{DS} because recrossings would lead to an overestimation of the rate otherwise.
The major importance of the \ac{TST}
follows from the broad variety of fields, where it can be applied to, 
including for instance
atomic physics~\cite{Jaffe00},
solid state physics~\cite{Jacucci1984},
cluster formation~\cite{Komatsuzaki99, Komatsuzaki02},
diffusion dynamics~\cite{toller, voter02b},
cosmology~\cite{Oliveira02}, 
celestial mechanics~\cite{Jaffe02, Waalkens2005b},
and Bose-Einstein condensates~\cite{Huepe1999, Huepe2003, Junginger2012a, 
Junginger2012b, Junginger2013b},
to name a few.
For \emph{time-dependent} systems, \eg~subject to periodical driving
due to external fields, the situation becomes more challenging.
Here, the \ac{DS} itself becomes time-dependent
and depends non-trivially on the saddle of the potential.
The \ac{DS} can nevertheless be obtained, e.g., by using a
minimization procedure based on the \acp{LD}
\cite{Mancho2010,Mancho2013,hern17h,hern19a}.

In a system with $d$ degrees of freedom, the time-dependent \ac{DS}
embedded in phase space has dimension $2d-1$ and is attached to the
$(2d-2)$-dimensional \ac{NHIM}, which has the property that every
trajectory on the \ac{NHIM} will never leave this manifold.
In a periodically driven system, we can construct and define a 
\ac{TS} trajectory which never escapes from the reactant region
and which is a periodic orbit with the same period as 
the driving potential \cite{dawn05a,dawn05b,hern06d,Kawai2009a,hern14b,hern14f,hern15a,
hern16a,hern16h,hern16i}.
In the limit of a system with one degree of freedom, the
\ac{NHIM} reduces to a point moving along the \ac{TS} trajectory.
However, in systems with two or more degrees of freedom the 
structure of the \ac{NHIM} and the dynamics
of trajectories on it becomes non-trivial.
Slowly reacting particles spend a longer time in the vicinity of the
\ac{NHIM} crossing the \ac{DS} closer to the \ac{NHIM}, and
therefore the dynamics on the \ac{NHIM} is of special interest
\cite{DynamicalReactionTheory2011}.

In this paper, we focus on the 
{numerical determination of the}
dynamics on the
\ac{NHIM} of a periodically driven system
{with more than one degree of freedom.}
Because of the unstable degree of freedom, trajectories on the
\ac{NHIM} tend to separate exponentially fast from this manifold.
This makes it difficult, if not impossible, to investigate numerically the long-time
behavior of the dynamics.
Here, we present a method that prevents numerically determined
trajectories from leaving the \ac{NHIM} by approximating the \ac{NHIM}
with a \ac{NN} and stabilizing the dynamics onto the \ac{NHIM}.
\acp{NN} have already found use in molecular dynamics \cite{wiggins17},
and have been seen to be  a powerful tool in the computation of
potential energy surfaces
\cite{blank1995neural,behler2007generalized,behler2011atom,
cui2016efficient,vargas2017machine,rupp2012fast,cui2015gaussian_a,
cui2015gaussian_b,faber2016machine,huang2017chemical} and the
construction of the \acp{DS} \cite{hern18c,hern19a}.

The challenge, addressed in this paper,
 is the determination of the 
 dynamics of trajectories 
within the 
  NHIM for systems with two or more degrees of freedom.
This is challenging because 
 trajectories in the neighborhood of the NHIM are unstable
  and depart exponentially fast from it.
  Applying \acp{NN} to stabilize the trajectories on the NHIM
  allows us to analyze them numerically using the tools
of nonlinear dynamics and thereby determine dynamical properties
such as reaction rates.
To illustrate the former, the dynamics on the \ac{NHIM} 
has been resolved in this work
using a stroboscopic
\ac{PSOS}.

For a periodically driven model system with two degrees of freedom we
show that the dynamics on the \ac{NHIM} is governed by torus-like
structures with a fixed point at its center representing a periodic
orbit with the same period as the driving potential.
We define this orbit as a periodic
\ac{TS} trajectory in analogy to systems with
one degree of freedom.
We have developed two methods, a centroid and a friction search
algorithm, to numerically extract such a
\ac{TS} trajectory in multidimensional systems.

Of particular importance is the question as to whether rate constants of
chemical reactions can be manipulated by periodic driving of the
system.
Rate constants can be obtained from appropriately chosen ensembles of
trajectories by evaluating the time-dependent number of reactive
trajectories having crossed the \ac{DS} \cite{hern17h,hern19a}.
In \REF~\onlinecite{hern14f} a Floquet analysis is used instead of ensemble
propagation to obtain the rate constant in a one-dimensional,
periodically driven model system with a moving saddle.
We generalize and apply the Floquet analysis to a periodic \ac{TS} trajectory
of the model system with two degrees of freedom
{obtained with either the centroid search or the friction search}
and show that the rate
constants computed with this method are in excellent agreement with
numerically much more expensive ensemble calculations.

Thus the central results of this paper are 
(i) the
demonstration of \acp{NN} for determining the multidimensional 
---viz in 2 dimensions--- dynamics of the NHIM,  and
(ii) the demonstration that detailed
  knowledge of the dynamics on the \ac{NHIM}, stabilized by \acp{NN},
  can be used to extend the
  one-dimensional Floquet method introduced 
in \REF~\onlinecite{hern14f} 
to obtain rate constants in systems
 with two or more degrees of freedom.

The paper is organized as follows.
In Sec.~\ref{sec:Theory and Methods} we introduce the theory and
methods to apply a \ac{NN} for the construction of the time-dependent
\ac{NHIM}, the stabilization of the dynamics on the \ac{NHIM}, and the
computation of rate constants.
In Sec.~\ref{sec:Results} results demonstrating the efficacy and
efficiency  of \acp{NN} for determining rate constants
are presented and discussed.
We conclude in 
Sec.~\ref{sec:Conclusion} that our new approaches do indeed provide
accurate and more efficient determinations of the rates of driven reactions
at increasingly higher dimensionality.

\section{Theory and Methods}
\label{sec:Theory and Methods}
In Sec.~\ref{sec:NHIM} we start with the definition and physical
fundamentals of the \ac{NHIM} including its numerical construction, and
give a short overview of \acp{NN}, used in Sec.~\ref{sec:stabilization}
to stabilize the dynamics on the \ac{NHIM}. 
In Sec.~\ref{sec:rate_constants} we then discuss the Floquet analysis
for the \ac{TS} trajectory and its application to determine rate
constants of the system with significantly reduced numerical effort
compared to methods based on the propagation of trajectory ensembles.

\subsection{Construction of the time-dependent \ac{NHIM}}
\label{sec:NHIM}
We consider a system with $d$ degrees of freedom, where the reactants
and products are separated by a time-dependent potential barrier
given by a rank-1 saddle.
A particle reacts, when it overcomes the saddle from one basin to the
other along the reaction coordinate $x$. 
The remaining $d-1$
coordinates $\vb{y}$ are called bath coordinates \cite{hern19a}.
The reaction pathways and the corresponding reaction rate of the
system are determined by the local properties of the barrier.
The \acf{NHIM} is the $(2d-2)$-dimensional manifold in the
$(2d)$-dimensional phase space which contains all trajectories that
are trapped in the saddle region both forward and backward in time.
Since we consider time-dependent systems, the \ac{NHIM} is in general
also time-dependent.
The dynamics on the \ac{NHIM} is of special interest for the reaction
dynamics, for the reason that the slower a particle reacts, the more
it is influenced by the saddle and the closer it passes the \ac{NHIM}.

A prerequisite for the investigation of the dynamics on the \ac{NHIM}
is the numerical construction and efficient description of this
manifold.
This aim can be achieved in two steps:
In a first step, we compute 
{a small number of} individual points
\begin{align}
  x^{\mathrm{NHIM}} &= x^{\mathrm{NHIM}}(\vb{y},\vb{v}_y,t) \; ,\nonumber \\
  v_x^{\mathrm{NHIM}} &= v_x^{\mathrm{NHIM}}(\vb{y},\vb{v}_y,t)
\label{eq:xv_NHIM}
\end{align}
on the \ac{NHIM} depending on the bath coordinates, velocities, and
time. Each of these points corresponds
to a point on one of the bound trajectories to the \ac{NHIM},
captured at a particular time $t$.
The stable and unstable manifolds $\Wsu$ play an important role in
this picture.
Here, the stable manifold $\Ws$ is the set of points that
approach the \ac{NHIM} exponentially fast, and the unstable
manifold $\Wu$ consists of those points that depart exponentially fast
from the \ac{NHIM}.
The points $(x^{\mathrm{NHIM}},v_x^{\mathrm{NHIM}})$ in Eq.~\eqref{eq:xv_NHIM} 
are given as the intersection of the stable and unstable manifolds
$\Wsu$, and can be numerically determined by application of the binary
contraction method \cite{hern18g,hern19a}.
{This two-dimensional bisection method
can be used to obtain
individual points on the \ac{NHIM} to high precision. 
However, this comes at the cost of computation time 
 because it requires the propagation of a large number of trajectories.}

In a second step, the high-accuracy points obtained in the first step
are used for the training of a \acf{NN}
as discussed in \REFS~\onlinecite{hern18c,hern19a} that allows for the
numerically fast and efficient interpolation to
an effectively continuous
set of
points on the \ac{NHIM}.
Any candidate trajectory can then be analyzed relative to the NN NHIM
to determine if it is reactive.
It can also be used to determine rates directly as shown below.

\begin{figure}
   \includegraphics[width=\columnwidth]{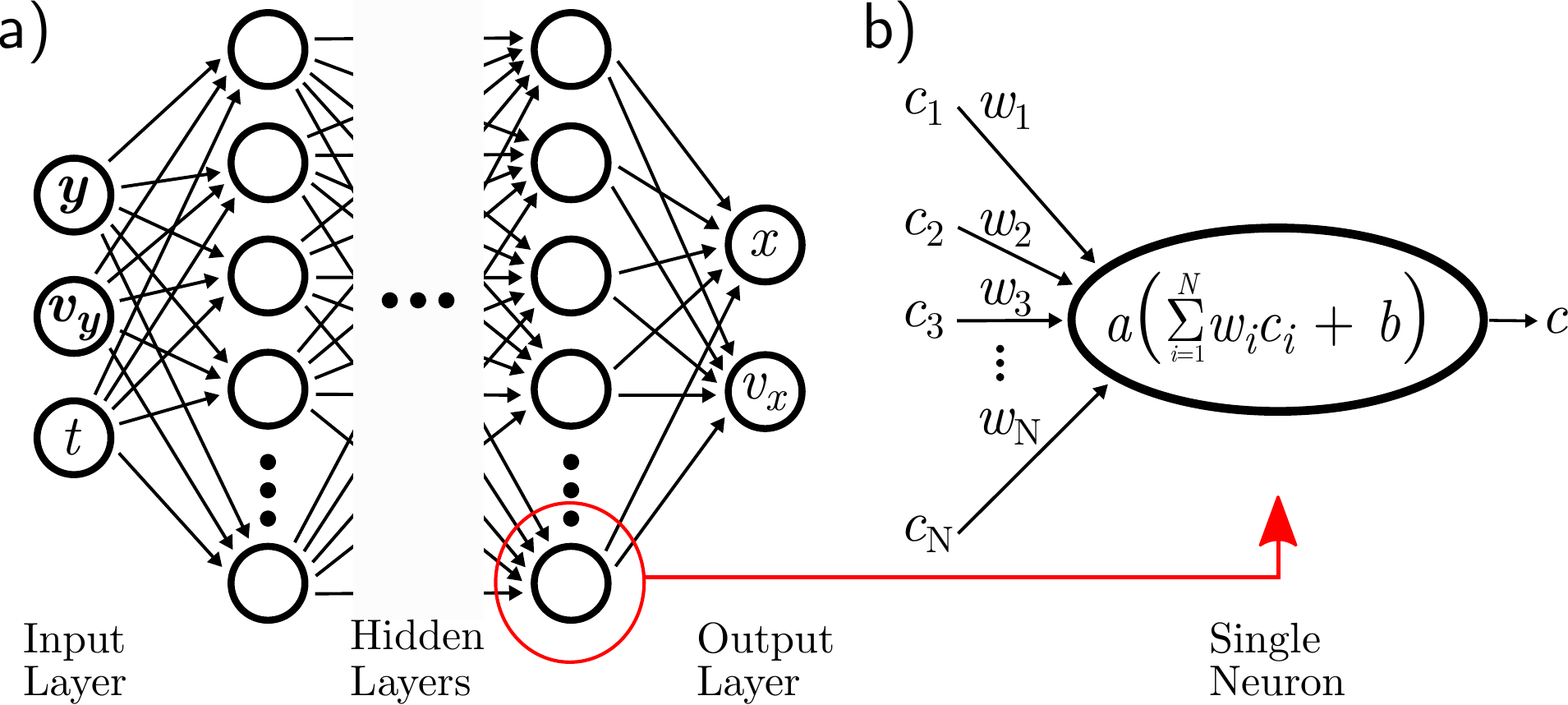}
    \caption{Basic construction of a feed-forward neural network~(a) and an
    individual neuron~(b).
    Here, we use such networks for the multidimensional regression task of
    interpolating the $(2d-2)$-dimensional NHIM of a $(2d)$-dimensional,
    time-dependent system. For any given time $t$, the reaction coordinates
    $(x, v_x)$ are continuously obtained using the network
    with the $(2d-2)$ phase space coordinates
    $(\vb{y}, \vb{v}_y)$ and the time $t$ as input. The weights $w$ and
    the biases $b$ are obtained using the \emph{Adam optimizer} (see text).
  }
  \label{fig:neural_net}
\end{figure}

For the convenience of the reader, we here briefly recapitulate the
basic ideas of \acp{NN}.
The more interested reader is referred to the literature,
{\it e.g.,} 
\REFS~\onlinecite{Goodfellow-et-al-2016,Nielsen-2015}.
In general, a \ac{NN} consists of \emph{layers} which can each
be represented as a
vector in a mathematical sense, see Fig.~\ref{fig:neural_net}(a).
Every layer is composed of \emph{neurons}, which correspond to the
entries of the vector and are represented in Fig.~\ref{fig:neural_net}(b).
A simple form of \acp{NN} is a \emph{feed-forward} neural net in which
information propagates only in one direction. 
The values for the $(l+1)$-st layer are
obtained as follows:
\begin{enumerate}
  \item Multiply the $l$-th \emph{weight matrix} $w_{ij}$ with the 
  values of the $l$-th layer  $c_j^{(l)}$.
  \item Add a \emph{bias vector} $b_{i}$ to the result of step 1.
  \item Apply the (usually) nonlinear \emph{activation function} $a(x)$.
\end{enumerate}
Mathematically, this corresponds to
\begin{equation}
  c_i^{(l+1)} = a \qty[ \qty(\sum_{j=1}^N
  w^{(l)}_{ij} c_j^{(l)} ) + b_i^{(l)} ] \,.
  \label{eq:NN-weighted-input-def}
\end{equation}
This is done layer by layer, starting from the input layer and ending
in the output layer.
For simplicity,
every neuron in a layer is assigned
the same activation function.
Although this limits the generality in the function space accessible
to the \ac{NN},
it increases the numerical stability of the optimizer.
In the present case, we chose
$a(z)=\tanh(z)$ for some input $z$
for all nodes except those in
the last layer for which a linear activation function is required.

The so called \emph{loss} or \emph{cost function} gives a measure of
the quality of the \ac{NN}.
For simplicity, we contract the multiple input features and
output labels into an
input vector $\vb{\gamma^\mathrm{i}}\equiv(\vb{y}, \vb{v_y}, t)^{\mathrm{T}}$
and an output vector
$\tilde{\vb{\gamma}}^{\mathrm{o}}(\vb{\gamma}^{\mathrm{i}})\equiv(\tilde{x}, \tilde{v_x})^{\mathrm{T} }$, respectively.
We use the mean squared error as the loss function
\begin{align}
  C_{\vb{w}, \vb{b}}(\vb{\gamma}^{\mathrm{o}},\tilde{\vb{\gamma}}^{\mathrm{o}}) = 
				\frac{1}{2n}\sum\limits_{i=1}^{n}\norm{ \vb{\gamma}^{\mathrm{o}}_i - 
				\tilde{\vb{\gamma}}^{\mathrm{o}}(\vb{\gamma}^{\mathrm{i}}_i) }^2 \,,
  \label{eq_NN_cost_function}
\end{align}
for the $n$ examples in the
\emph{test data} or \emph{training data}
of points in space and time of the NHIM.
The data set
$\{\vb{\gamma}^{\mathrm{i}}_i,\vb{\gamma}^{\mathrm{o}}_i\}$ with $i\in[1, n]$
is obtained with the binary contraction method mentioned above
and described in \REFS~\onlinecite{hern18g,hern19a}.
The 
subscripts $\vb{w}, \vb{b}$ indicate the dependence on the weight
matrices and the bias vectors.

During the training of the \ac{NN} the weights, i.e., the values of the
weight matrices and the bias vectors, are adjusted such that the loss
given in \EQ~\eqref{eq_NN_cost_function} is minimized.
In our case, this is done by
a modified version of \emph{stochastic gradient descent}~\cite{robbins1951sgd},
which is called an \emph{Adam optimizer}~\cite{diederik2014optimizer}.
{The learning rate $\eta$, determining the step size of the gradient descent method,
is set to $\eta = 0.1$ and all other optimizer parameters
are in accordance to the original publication~\cite{diederik2014optimizer}.}
In comparison to previous
works like \REF~\onlinecite{hern18c}, this simplifies the choice of
hyper-parameters. 
The \ac{NN} and its optimization is implemented
using the \emph{Python} library \emph{Tensorflow}~\cite{tensorflow2015-whitepaper}.
{All \acp{NN} implemented here have
three neurons in the input layer,
two neurons in the output layer, and 
three hidden layers with 100, 100, and 40 neurons,
  respectively in order.}

An aim of this paper is to use \ac{NN}s to 
effect a multidimensional regression task. 
In particular, for $d$-dimensional systems we need a 
continuous description of the $(2d-2)$-dimensional
\ac{NHIM} that is embedded in the $(2d+1)$-dimensional
\emph{extended phase space} (phase space plus time).
We use $2d-2$ phase space coordinates $\vb{y}$,
$\vb{v}_y$ and the time $t$ as input vectors and obtain the
corresponding two phase space coordinates $(x, v_x)$ on the \ac{NHIM},
see Eq.~\eqref{eq:xv_NHIM} and Fig.~\ref{fig:neural_net}(a).
{The training data can be generated as described above using
  the binary contraction method \cite{hern18g, hern19a}.
  Training was done with $50\,000$ training points over $50\,000$
  epochs, which took about 2 hours of computational time on
  on an \emph{Intel(R) Core(TM) i5-3470} CPU with $3.20\,$GHz.}

{In comparison to the direct application of the binary contraction,
  the \ac{NN} takes a factor of 200 less in computational time.
  It suffers a reduction in precision from about $10^{-15}$
  (for the binary contraction method) to about $10^{-4}$ (for the \ac{NN}).
  However, the stabilization of the dynamics on the 
  \ac{NHIM} does not require very high precision, as will be discussed
  in Sec.~\ref{sub:Localizing TSt}.
Alternative machine learning methods like Gaussian Progress Regression
(GPR) could be used as well. However, due to the relative high number of
trainings points (about $50\,000$) that are needed for the required accuracy, 
inferring of GPR would be expected to be
slower than the \ac{NN} in this case.
More details are given in \REF~\onlinecite{hern19a}.}
We anticipate that future work could perhaps
obtain improved efficiency
by optimizing the \acp{NN} through advances in machine learning 
techniques. However, this
would not alter the physical interpretation of the present results.

\subsection{Stabilization of trajectories on the \ac{NHIM}}
\label{sec:stabilization}
Due to the unstable degree of freedom, trajectories tend to move away
from the \ac{NHIM}.
The distance between the orbit and the \ac{NHIM} increases
exponentially fast in time, and thus the initial conditions of a
trajectory must be known with numerically inaccessible precision to
keep it on the \ac{NHIM} for long times.
This makes a long-time analysis of the dynamics on the \ac{NHIM}
extremely difficult or even impossible.

Here, we present a method for stabilizing the motion of trajectories on
the \ac{NHIM}.
In Sec.~\ref{sec:NHIM} we have shown that a \ac{NN} can be used to
describe the $(2d-2)$-dimensional \ac{NHIM}.
Here we apply this \ac{NN} to guide the trajectory on the
unstable manifold, i.e., to correct any deviation from the \ac{NHIM}
during the numerical integration of the orbit.
The procedure is as follows: 
Using the \ac{NN}, we choose an arbitrary initial point on the \ac{NHIM}.
Note that the uncertainty of the initial conditions is determined
by the numerical accuracy of the \ac{NN}.
From this point, the trajectory is propagated for a small time step.
{This numerical} step may have {naively} increased the distance between the
trajectory and the \ac{NHIM}. 
We assume that this deviation is manifested mainly in the reaction
coordinate $x$ and the corresponding velocity $v_x$, indicating the
falloff of the trajectory from the \ac{NHIM} along the unstable
direction.
It is thus necessary to guide the
trajectory back to the \ac{NHIM}.
We achieve this by replacing the {calculated} 
position $x(t)$ of the reaction
coordinate and the corresponding velocity $v_x(t)$, which may slightly
deviate from the \ac{NHIM}, with the (numerically) exact values,
{$x$ and $v_x$}, on the
\ac{NHIM} given by 
{the \ac{NN} based on the 
data set of points from Eq.~\eqref{eq:xv_NHIM}.}
{The error here is limited,
as discussed in
Sec.~\ref{sec:NHIM}, 
because the error of the \ac{NN} itself is small
compared to the exponential increase of the deviation along the
unstable reaction coordinate in time without the stabilization.}
By repeating this procedure after each time step, we prevent the
trajectory from leaving the \ac{NHIM}.
{Note that this procedure is not limited to systems with two
degrees of freedom and can be applied for any dimension of
the problem as long as the data base feeding the NN sufficiently
spans the underlying domain of $(\vb{y}, \vb{v}_y , t)$.
Furthermore, this construction does not require (and is not restricted to)
either the trajectories or the external driving 
to be periodic in time as assumed in the illustrative examples below.}

For the special case of a periodically driven system with $d=2$
degrees of freedom, discussed in Sec.~\ref{sec:Results}, the
stabilization restricts the dynamics of the trajectories from the
$(4+1)$-dimensional extended phase space (including time) of
$H(x,v_x,y,v_y,t)$, as discussed in \REF~\onlinecite{hern19a},
to the $(2+1)$-dimensional subspace of the \ac{NHIM}, viz.
\begin{align}
  & H\left(x^{\mathrm{NHIM}}(y,v_y,t),v_x^{\mathrm{NHIM}}(y,v_y,t),y,v_y,t\right) \nonumber\\
   = & H^{\mathrm{NHIM}}(y,v_y,t) \, .
\end{align}
This is effectively a periodically driven one degree of freedom
system.
As such, it can in principle show regular, chaotic or mixed
regular-chaotic dynamics \cite{Breuer1990}.
The dynamics can be visualized using a \ac{PSOS} with a stroboscopic
map for time, i.e., points $(y(t), v_y(t))$ are drawn at times $t
= t_B + nT$ with $t_B$ the barrier phase, $T$ the period of the
driving, and $n\in\mathbb{N}$.
The analysis of the \ac{PSOS} allows one to deduce the dynamics of the
system from structures of the trajectories in the \ac{NHIM}.
In particular, periodic orbits
appear as fixed points.
{Linear structures in the \ac{PSOS} indicate the existence of
  exact or approximate constants of motion and thus regular dynamics.
  Chaotic dynamics is related to the breakdown of constants of motion
  and is visualized as stochastic regions in the \ac{PSOS}
  \cite{Lichtenberg82}.}

A fixed point with period one in the \ac{PSOS} indicates a periodic
orbit with the same period as the driving potential.
In analogy to systems with one degree of freedom
\cite{dawn05a,dawn05b,hern06d,Kawai2009a,hern14b,hern14f,hern15a,
hern16a,hern16h,hern16i}
we define this orbit as the \ac{TS} trajectory.
It is of major importance for the following reasons.
In dissipative, non-periodic systems the \ac{TS} trajectory is
the unique trajectory that is bound to the vicinity of the saddle for
all times \cite{hern16i,dawn05a,dawn05b}.
In Hamiltonian systems, the \ac{TS} trajectory has, at least
qualitatively, only the minimal amount of the (in driven systems not
conserved) energy that is necessary to follow the saddle motion and
the least energy for motion in additional directions.
It is therefore located in the low energy regime of the barrier.
This holds also for trajectories in dissipative systems, since their
additional energy is damped with time.
Trajectories in the low energy regime of the barrier are assumed to be
strongly influenced by the saddle.
On the other hand, the \ac{TS} trajectory allows for the calculation
of the reaction rate by an alternative method, to the often used
ensemble propagation \cite{hern14f,hern15a}.

\subsection{Rate constants}
\label{sec:rate_constants}
As mentioned above, the \ac{NHIM} is of particular importance because
the dynamics close to this manifold determines the rate constants of
trajectories crossing the saddle region.
Rate constants can be computed by 
propagating a large ensemble of trajectories 
starting close to the \ac{DS} and subsequently fitting the number of
trajectories $N_{\mathrm{react}}(t)$ that remain reactants over time
before crossing the moving \ac{DS} to the exponential form
\begin{equation}
  N_{\mathrm{react}}(t) - N_\infty \propto \exp(-kt)
\end{equation}
where $k$ is the resulting rate constant.
Details are given in \REFS~\onlinecite{hern17h,hern19a}.

For systems with one degree of freedom, a computationally less
expensive method for obtaining rate constants has been derived by
Craven et al.~\cite{hern14f}.
Here, we give a short review of this method, which is based on the
Floquet analysis of the \ac{TS} trajectory.

For simplicity we generalize the configuration space coordinates $x_i$
(for the $i$-th degrees of freedom, in a system with $d$ degrees of freedom)
and the momentum coordinates $p_i$ to the phase space coordinates
\begin{equation}
  \vb{\gamma} = (x_1,x_2,\dots,x_d,p_1,\dots,p_d)^{\mathsf{T}}\,.
  \label{eq:Floquet_phase_space_vector}
\end{equation}
With this notation the equations of motion transform to
\begin{equation}
  \dot{\vb{\gamma}} = \vb{J} \pdv{H}{\vb{\gamma}}\quad \,
  \textrm{with} \quad \vb{J}=
  \begin{pmatrix}
    \vb{0}_d & \vb{1}_d \\
    -\vb{1}_d & \vb{0}_d 
  \end{pmatrix}
  \,,
  \label{eq:Floquet_hamiltonian}
\end{equation}
where $H$ denotes the Hamiltonian of the system, and $\vb{1}_d$ and
$\vb{0}_d$ describe the $d$-dimensional identity and zero matrix,
respectively.
Based on this, the \emph{stability} or \emph{monodromy matrix} for a
trajectory starting at $\vb{\gamma}(0)$ is defined according to
\cite{goldstein1980classical}
\begin{equation}
  \vb{M}_{ij} \qty[ \vb{\gamma}(0),t] = \pdv{\gamma_i(t)}{\gamma_j(0)} \, .
  \label{eq:Floquet_stability_matrix}
\end{equation}
By considering two initially neighboring trajectories, it follows by
chain rule from \EQS~\eqref{eq:Floquet_hamiltonian} and
\eqref{eq:Floquet_stability_matrix} the differential equation
\begin{equation}
  \dot{\vb{M}} = \vb{J} \pdv[2]{H}{\vb{\gamma}}\vb{M} \, , \quad
  \vb{M}(0) = \vb{1}_{2d} \, .
\label{eq:Floquet_differential_eq}
\end{equation}
For Hamiltonian systems, \ie~in systems without friction,
the monodromy matrix $\vb{M}$ is symplectic.
This means, if $\lambda$ is an eigenvalue of $\vb{M}$ then
$1/\lambda$ and their complex conjugates $\bar{\lambda}$, 
$1/\bar{\lambda}$ are also eigenvalues.
Since the monodromy matrix has only real entries, the eigenvalues
are either complex conjugated or inverse to each other. 

In the following $m_{\mathrm{l}}$ denotes
the larger eigenvalue and $m_{\mathrm{s}} = 1/m_{\mathrm{l}}$
denotes the smaller one. For periodic trajectories with period $T$, 
the so called \emph{Floquet exponents} are defined by
\begin{equation}
  \mu_{\mathrm{l,s}} = \frac{1}{T} \ln \abs{m_{\mathrm{l,s}}} \; .
\end{equation}
An important fact about the Floquet exponents is that they correspond
to the rate at which two neighboring trajectories separate from each other
\cite{goldstein1980classical}.
Eigenvalues with absolute value equal to one correspond to vanishing
Floquet exponents.
Hence, neighboring trajectories do not separate exponentially.
This corresponds to a stable degree of freedom.
On the other hand, eigenvalues with absolute value unequal to one
correspond to non-vanishing Floquet exponents and therefore to an
unstable degree of freedom.
Thus, in a one-dimensional system the reaction rate constant is determined
by the two Floquet exponents of the \ac{TS} trajectory and follows
from \REF~\onlinecite{hern14f} as
\begin{equation}
  \label{eq:Reaction_rate_from_Floquet_exponents}
  k_{\mathrm{Floquet}} = \mu_{\mathrm{l}} - \mu_{\mathrm{s}} \; .
\end{equation}

\section{Results and discussion}
\label{sec:Results}
We now benchmark the methods described above using
a system with non-trivial dynamics---\ie~one that contains a mix
of regular and chaotic behavior~\cite{Breuer1990}---on a time-dependent \acs{NHIM}.
Such behavior arises readily in nonlinear coupled
multidimensional systems with high dimensionality.
Indeed, most chemical reactions have such a structure.
Nevertheless, for relative simplicity, we consider a low-dimensional
time-dependent model reaction which admits detailed analysis
while retaining the requisite complexity.
Further requirements to the model system are the existence of a rank-1 saddle and 
a periodic oscillation in the saddle's position.

\subsection{Model System}\label{sec:model_system}

The two-dimensional model system from \REF~\onlinecite{hern17h} with the
time-dependent potential
\begin{align}
  V(x, y, t) & =
  E\sno{b}\,\exp\left(
  -\alpha\left[x- \hat{x} \sin\left(\omega_x t \right)\right]^2\right) \nonumber\\
  & + \frac{\omega_y^2}{2}\left[y-\frac{2}{\pi}
  \arctan\left(2 x \right)\right]^2
  \label{eq:potential}
\end{align}
satisfies our requirements for verifying the accuracy of the approach
through a nontrivial example.
It contains a Gaussian barrier with height $E_\mathrm{b}$ and width $a$,
and oscillates along the $x$ axis with frequency $\omega_x$
and amplitude $\hat{x}$. 
It is bound along the $y$ direction through a harmonic potential
with frequency $\omega_y$, and is nonlinearly coupled to
the minimum energy path
$x$ through a term proportional to $y\arctan(2x)$.
For simplicity, all variables are dimensionless
and in accordance to \REF~\onlinecite{hern17h} set to $E_\mathrm{b} = 2$, 
$\alpha = 1$, $\omega_x  = \pi$, and $\omega_y = 2$.
The amplitudes $\hat{x}$ of the 
oscillations in the saddle range between
$0$ and $0.8$.

\subsection{Dynamics on the \acs{NHIM}}
\label{sub:Localizing TSt}
\begin{figure}
  \includegraphics[width=0.9\columnwidth]{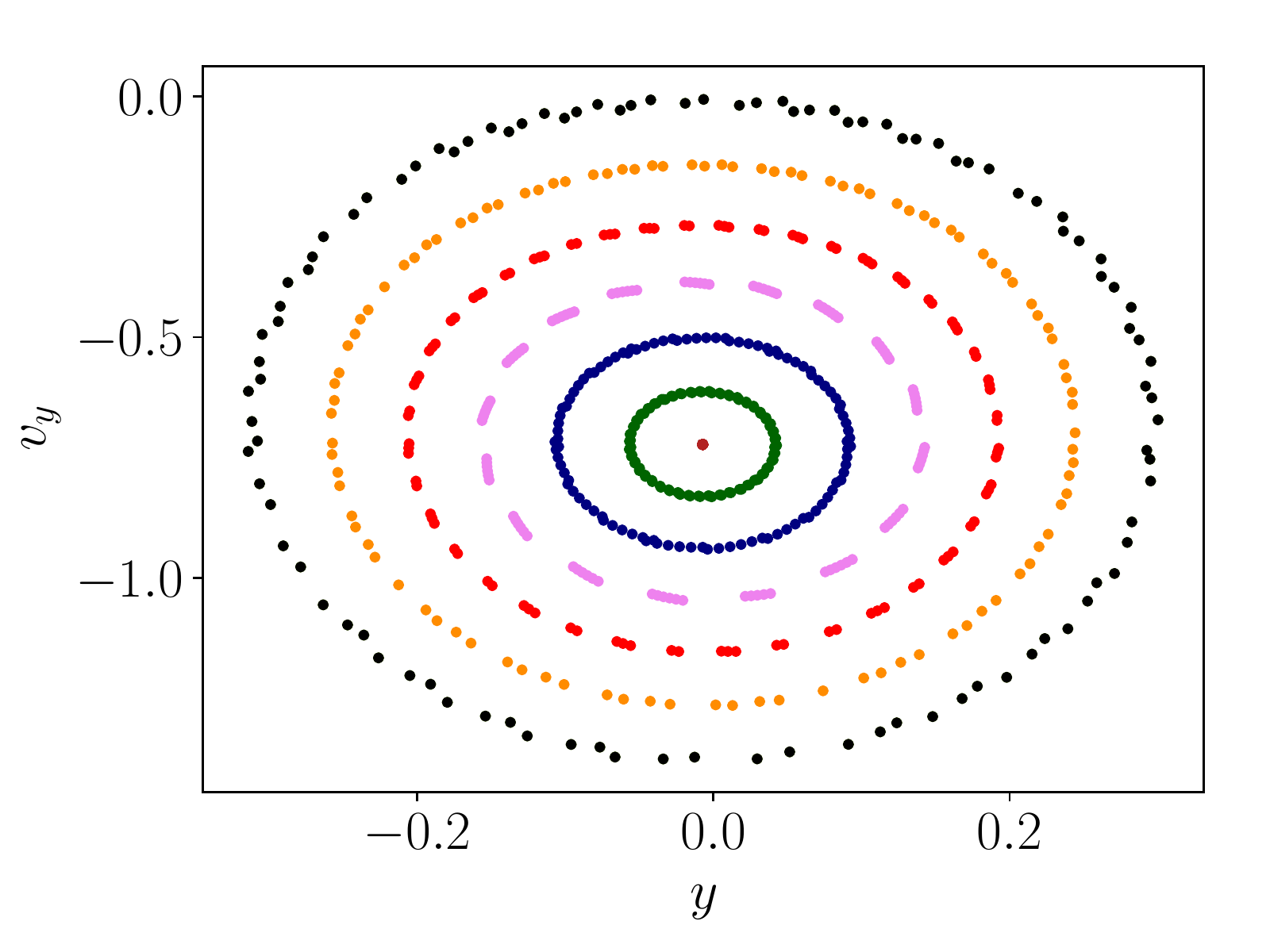}
  \caption{\acf{PSOS} of the system (\ref{eq:potential}) with
    amplitude $\hat x = 0.4$ using a stroboscopic map with barrier
    phase $t_B=0$.  Trajectories with various initial conditions have
    been stabilized on the \ac{NHIM} via a \ac{NN} and propagated for
    100 periods of the driving potential.  The fixed point at the
    center of the regular torus-like structures marks the \ac{TS}
    trajectory.}
  \label{fig:PSOS}
\end{figure}

To visualize trajectories we use Poincar\'{e} surfaces of
section as trajectories cut through the $y$ and $v_y$ plane.
This works well in the current case of a 
time-dependent and periodically driven system because the
response of the system is similarly periodic.
Consequently, a stroboscopic representation taken at
intervals matching the period of the driving,
such as that shown in \FIG~\ref{fig:PSOS},
provides a view of stable structures, such as the tori,
or unstable ones, if they appear in the system.

The trajectories shown in the stroboscopic view 
in \FIG~\ref{fig:PSOS} 
were sampled around the \ac{TS} trajectory, which
is obtained by an algorithm that is described later.
It is remarkable that all trajectories approximately lie on ellipses,
meaning that the phase space contains stable tori.
This implies that there is an approximately conserved quantity
for a fixed barrier phase~\cite{Wimberger2014}.
In other words, the system shows regular behavior.
It is robust to changes in the 
oscillation amplitude as we found it
to persist at various values, {\it i.e.}, $\hat{x} = 0.0,\, 0.1$, or $0.8$.
Consequently, these systems are near-integrable in the regime close to
the rank-1 saddle. This does not necessarily hold for regimes far from the saddle
or for arbitrary system parameters.

As mentioned before, we want to localize the \ac{TS} trajectory in a
periodically driven system.
In the \ac{PSOS}, the \ac{TS} trajectory should appear as a
fixed point in the stroboscopic view because the periodicity of the trajectory
is equal to that of the driven barrier motion.
In the following, we present two algorithms for finding a point on the 
\ac{TS} trajectory from which the trajectory itself can be obtained 
by using the stabilization via a \ac{NN}.

\begin{figure}
  \includegraphics[width=0.9\columnwidth]{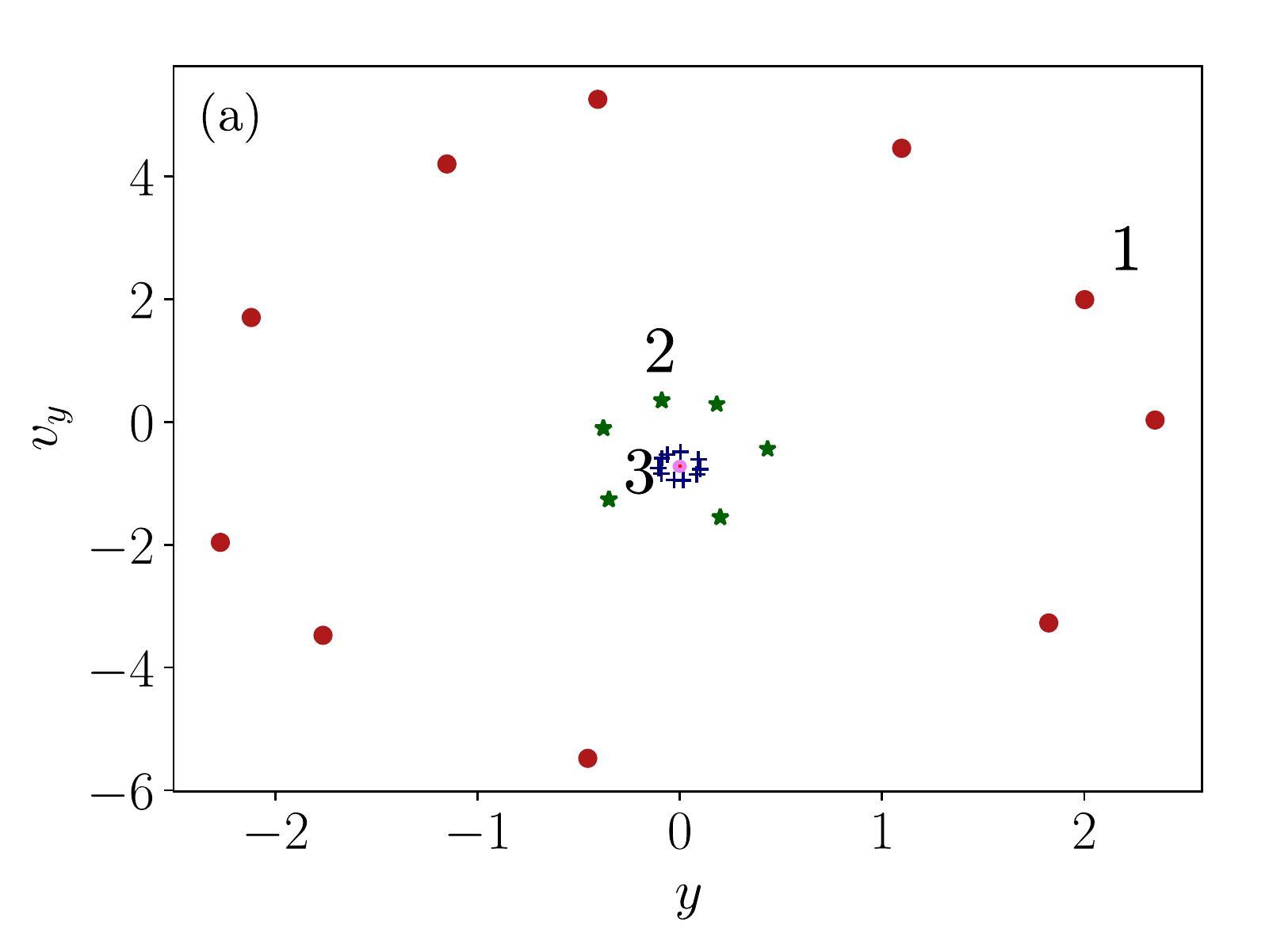}
  \includegraphics[width=0.9\columnwidth]{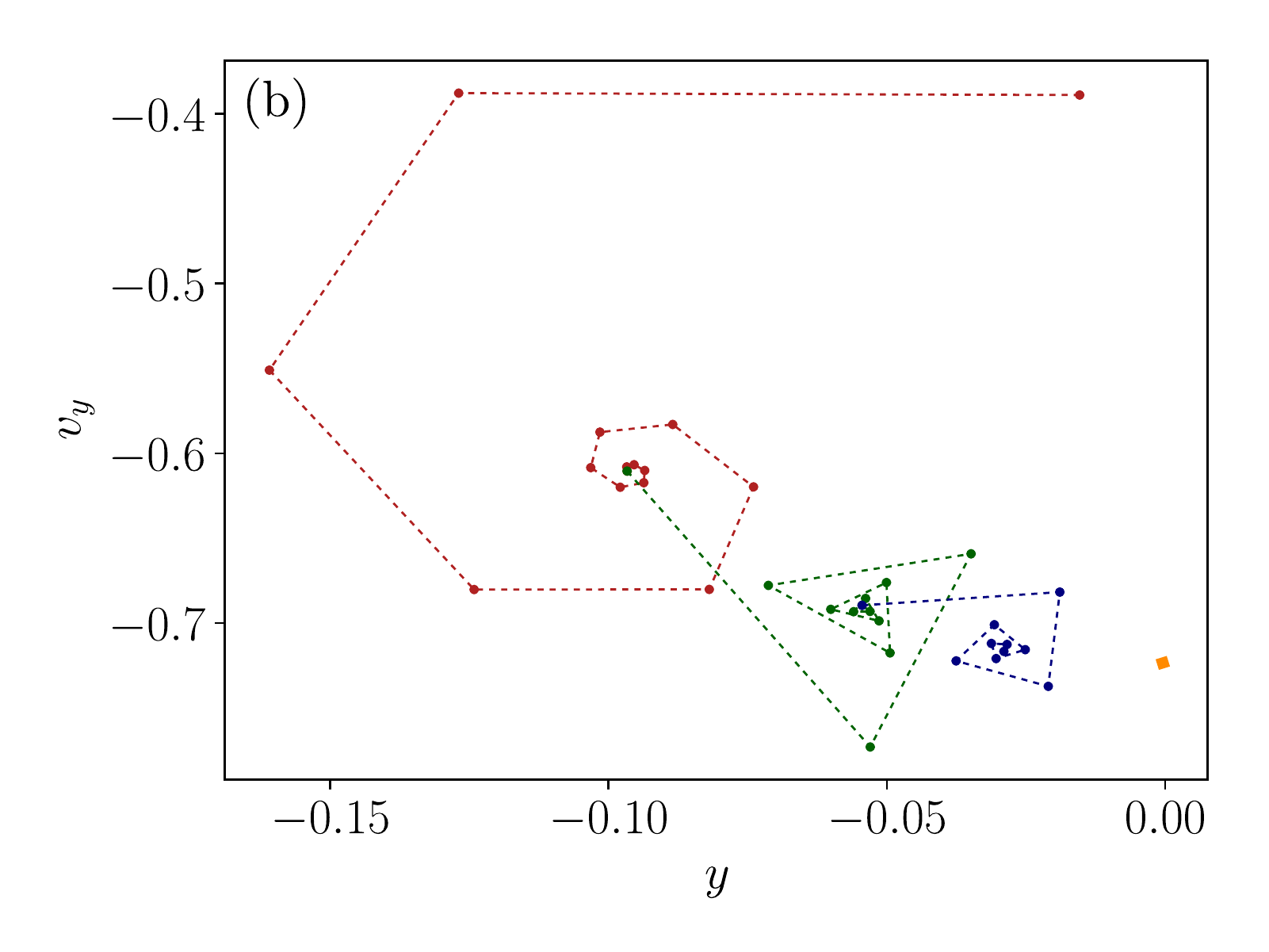}
  \caption{(a) Centroid search for the \ac{TS} trajectory of 
  system \eqref{eq:potential} with $\hat{x} = 0.4$.
    Stroboscopic view of trajectories for 10 periods determined 
    using stabilization generated with a \ac{NN}. 
Three selected initial points of trajectories are marked as 1, 2 and 3,
and correspond to the stroboscoped points marked as blue red dots, green stars
and blue crosses, respectively.
The centroid of the
    $n^{\mathrm{th}}$ trajectory corresponds to the initial point 
    of the $(n+1)^{\mathrm{st}}$ trajectory,
{\it e.g.} point 2 is the centroid of the
red points of trajectory 1.
The algorithm
    converges to the \ac{TS} trajectory, located, for the given
    barrier phase $t_B = 0$, at $y \approx 0.00$, $v_y \approx -0.72$.
    (b) Friction search for the \ac{TS} trajectory. The algorithm
    starts with large friction (red line) for the \ac{TS} trajectory
    starting point. The trajectory forms a spiral, starting in the 
    upper right corner and converging to a point that is closer
		to a periodic TS trajectory than the initial point.
    During the propagation of the particle the friction was
    reduced twice (green and blue line). 
    The initial point of the green curve corresponds to the
    final point of the red curve.
    The orange dot marks the
    point to which the friction search finally converges {by successively
    reducing the friction}.  Details of
    both algorithms are given in the text.}
    \label{fig:TS-trajectory_search}
\end{figure}
The first algorithm, which we call \emph{centroid search} uses the fact,
that we have a periodic system, so the potential
is identical after a full period in time has passed. Note that the \ac{PSOS}
in \FIG~\ref{fig:TS-trajectory_search} is given for a fixed barrier phase
$t_B = 0$.
The result is shown in a stroboscopic view in
\FIG~\ref{fig:TS-trajectory_search}(a).
The iterations of the first initial point, marked with a ``1'' lead to the
red dots.
The centroid or geometric center of these points is the green asterisk
marked with a ``2'' and is the initial point of the next trajectory.
The next iteration yields the initial point ``3'' and the blue plus
symbols.
As can be seen, the algorithm converges rapidly to a fixed point for
the particular barrier phase plotted.
This means that the trajectory period is equal to the system period.
For all investigated initial coordinates, the algorithm converged to
the same trajectory which is the only one that 
has a periodicity equal to the system periodicity. Therefore it is in 
accordance to the definition in Sec.~\ref{sec:stabilization}
and must be a periodic \ac{TS} trajectory.
\FIG~\ref{fig:TS-trajectory_search}(a) also shows that the centroid
search converges within a small number of iterations
and is thereby computationally inexpensive.

The centroid search algorithm is restricted to periodically driven
systems because it requires a stroboscopic view.
To overcome this problem, we developed a second algorithm.
We will once again take advantage of the 
\ac{TS} trajectory which has thus far been seen in this work as a fixed point of
the stroboscopic map in the context of 
periodic systems. 
In non-periodic systems, a \ac{TS} trajectory is known to be
a trajectory bound to the vicinity of the saddle for all time.

The second algorithm, which we call \emph{friction search}, 
relies on the introduction of an auxiliary
friction that reduces the energy of an arbitrary trajectory on the \ac{NHIM}.
The latter converges to that specific trajectory with the lowest
possible energy on the \ac{NHIM}.
High friction in the algorithm leads to a fast convergence,
since the energy dissipation is high, but the error in determining 
the \ac{TS} trajectory is large. This can be seen in 
\FIG~\ref{fig:TS-trajectory_search}(b).
Low friction decreases the convergence speed but increases the
precision.
Therefore we start with a large friction {at an arbitrary point} 
(red curve in \FIG~\ref{fig:TS-trajectory_search}(b)) and {propagate
until the trajectory converges. Then the friction coefficient is decreased and the prior
convergence point is used as the new initial point of the propagation. 
Propagation and friction coefficient reduction is done}
iteratively
(green and blue curves in \FIG~\ref{fig:TS-trajectory_search}(b))
to get a precise \ac{TS} trajectory starting point.
\FIG~\ref{fig:TS-trajectory_search}(b) shows that the friction search
converges to a similar value as the centroid search, which is marked
by an orange dot.
The slow energy dissipation for small friction is the reason why 
friction search converges slower than the centroid search, however,
this second method should work reliably in systems with many degrees of freedom,
and for arbitrary driving.

\subsection{Analysis of the \ac{TS} trajectory and rate constants}
\label{sub:Analysis_for_the_TSt}
In this section we use the point on a periodic \ac{TS} trajectory determined by the centroid
search to obtain this \ac{TS} trajectory itself. 
We obtain the latter by 
using either a conventional numerical integrator 
or the stabilization on the \ac{NHIM} via an \ac{NN}. 
\begin{figure}
  \includegraphics[width=0.9\columnwidth]{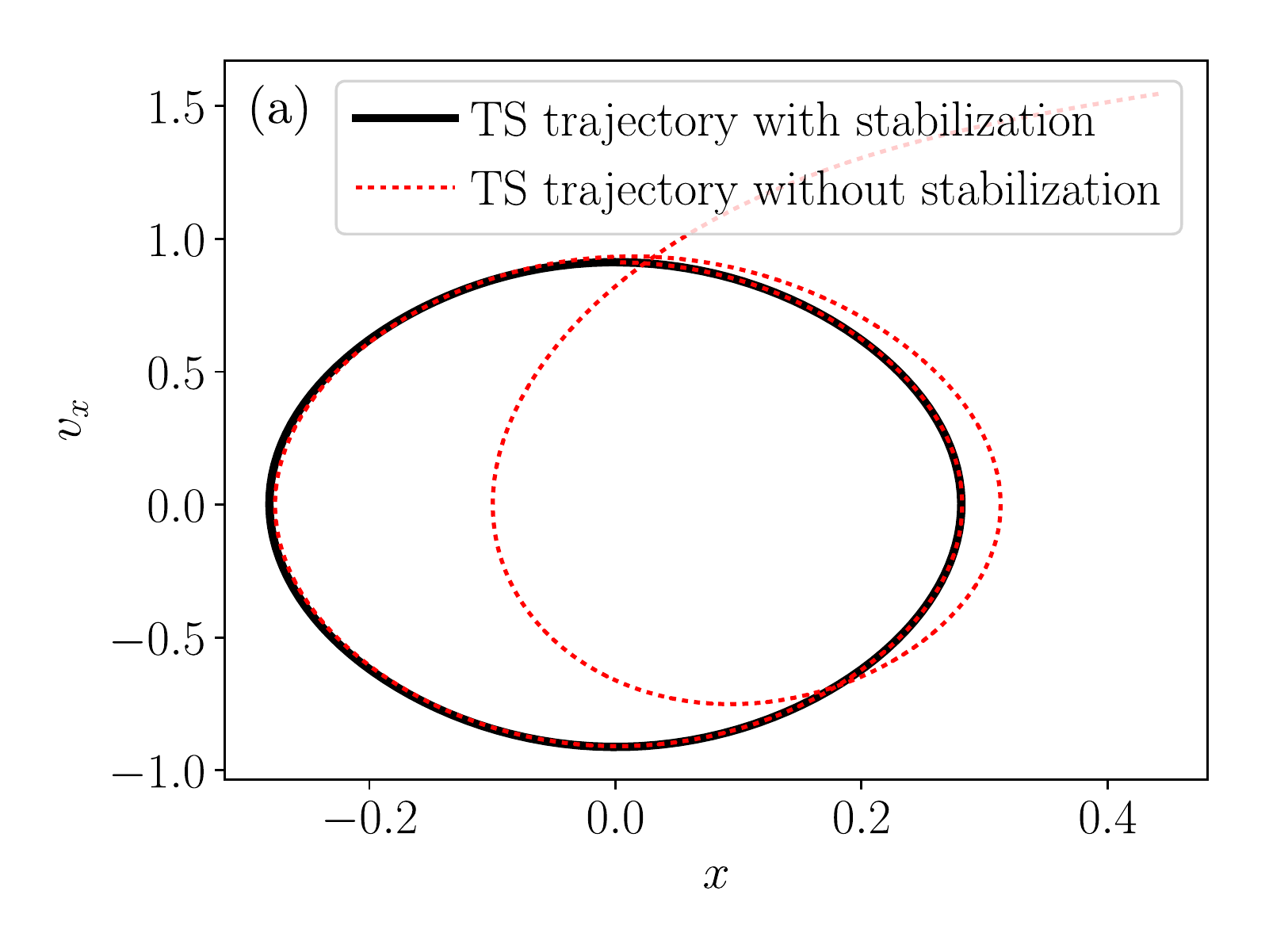}
  \includegraphics[width=0.9\columnwidth]{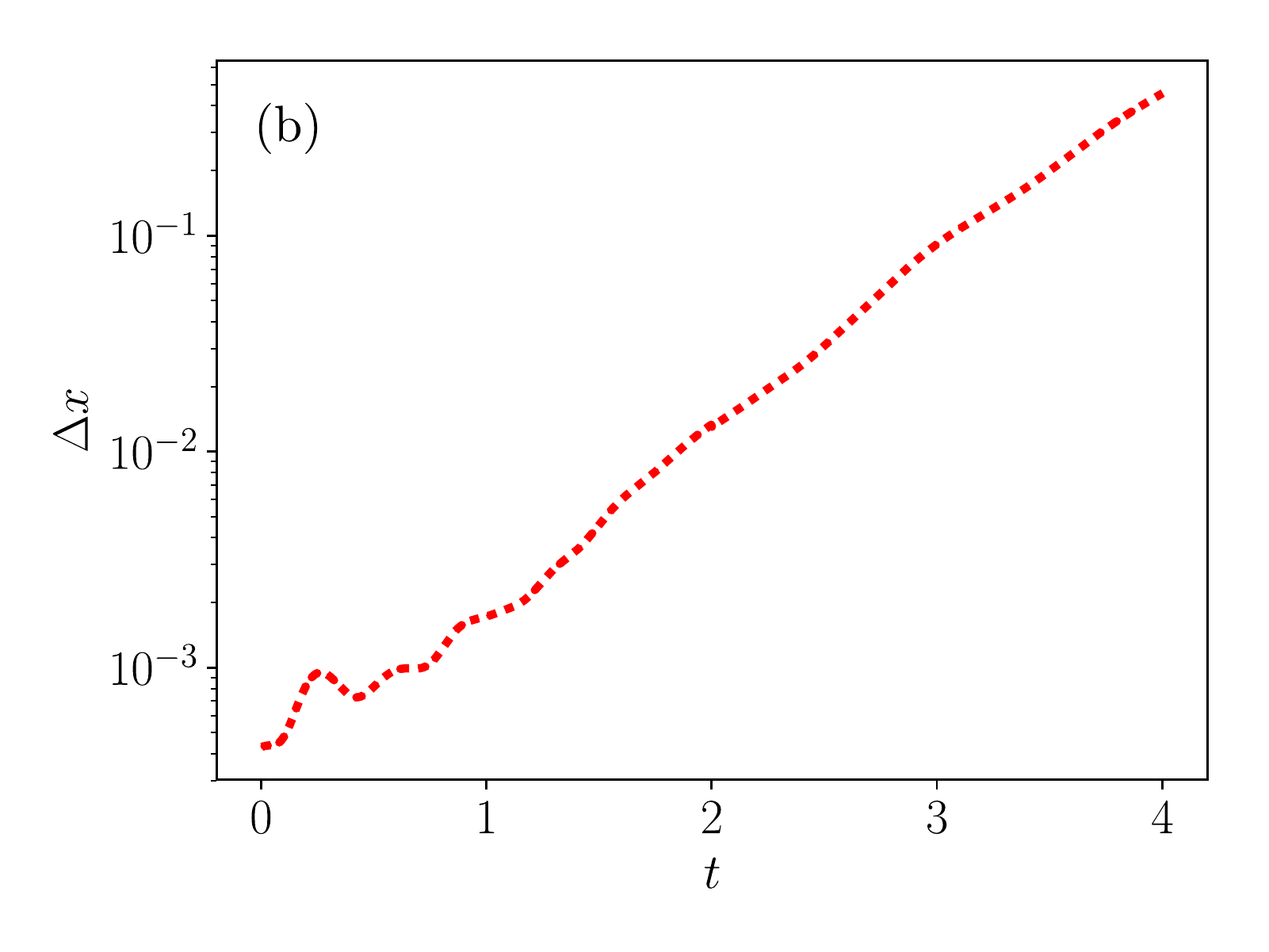}
    \caption{
    (a) Projection of the \ac{TS} trajectory, determined
    by the centroid search. Trajectories are plotted over $2.5$ periods 
    for both, the trajectory determination with and without the stabilization
    on the \ac{NHIM}.
    (b) Distance $\Delta x$ between the $x$-coordinates of the two
    trajectories calculated with and without the stabilization over
    two periods.  Within one period the separation is two orders of
    magnitudes smaller than the actual motion in $x$ direction.}
    \label{fig:TSt}
\end{figure}
The results are shown through 
a non stroboscopic view in \FIG~\ref{fig:TSt}(a).
There is great agreement between the two
different trajectories during at least the first period.
The distance in the $x$-direction
between the two trajectories reported 
in \FIG~\ref{fig:TSt}(b) shows nearly exponential separation.
This is due to the fact that the unstable degree of freedom of the 
rank-1 saddle causes the trajectory,
which is close to the \ac{NHIM} but not directly on the \ac{NHIM},
to increasingly deviate from the latter.
Significantly, the agreement in the trajectories during the first period
seen in \FIG~\ref{fig:TSt}b confirms that
the modification
(i) does not change the underlying physical behavior of the trajectory
and (ii) prevents the trajectory from leaving its perfect
periodic orbit.

\begin{table}
  \caption{Eigenvalues $m$ and Floquet exponents $\mu$ of the
				periodic \ac{TS} trajectory
  (determined with centroid search) for the potential barrier
  amplitude $\hat{x} = 0.4$.}
  \label{tab:Floquet}
  \centering
  \begin{tabular}{*2c}
    \toprule
    $m$ 				& $\mu$ 	\\ 
    \midrule
    $44.98$ 			& $+1.903$  \\ 
    $0.02223$ 			& $-1.903$  \\ 
    $0.4373 + 0.8993i$ 	& $0$  		\\ 
    $0.4373 - 0.8993i$ 	& $0$  		\\ 
    \bottomrule
  \end{tabular}
  \par
\end{table}

The stabilized trajectory (black line in Fig.~\ref{fig:TSt}(a)) is
used for the Floquet analysis in the following.
Table~\ref{tab:Floquet} shows the eigenvalues $m$ of the monodromy
matrix and Floquet exponents $\mu = (1/T) \ln \abs{m}$ (with $T$
describing the period time) of the periodic \ac{TS} trajectory.
It can be seen that the two real eigenvalues are inverse to each other
and the two complex ones are complex conjugated. This shows that
the monodromy matrix of our \ac{TS} trajectory is indeed symplectic, which is
in agreement with the theory. 

\begin{table}
  \caption{
  Reaction rate constants $k$ obtained by Floquet analysis of the
  \ac{TS} trajectories and by ensemble propagation for different barrier
  oscillation amplitudes $\hat{x}$.}
  \label{tab_reaction_rates2}
  \centering
  \begin{tabular}{*5c}
    \toprule
    $\hat{x}$ \
    & $k_{\textrm{Floquet}}$\
    & $k_{\textrm{Ensemble}}$\\ 
    \midrule
    $0.0$ 	& $2.761$  & $2.762$   \\
    $0.1$ 	& $2.979$  & $3.020$   \\
    $0.4$ 	& $3.806$  & $3.804$   \\
    $0.8$ 	& $4.016$  & $3.994$   \\
    \bottomrule
  \end{tabular}
  \par
\end{table}

The aforementioned Floquet exponents yield via 
\EQ~\eqref{eq:Reaction_rate_from_Floquet_exponents} the corresponding 
reaction rate constant, e.g., $k_{\mathrm{Floquet}} = 3.806$ for the
model system (\ref{eq:potential}) with barrier amplitude $\hat{x} = 0.4$.
The same analysis is done for three other sets of system parameters 
($\hat{x} = {0.0}, {0.1}, {0.8}$) and the results are shown in
Table~\ref{tab_reaction_rates2}.
For comparison, the reaction rates were also computed by observing 
the population decay of an ensemble of trajectories, and the resulting
rates are listed in the second column of
Table~\ref{tab_reaction_rates2}.
The agreement is excellent.
It provides further verification of the earlier conjecture
of Craven et al.~\cite{hern15a} stated in 
Eq.~\eqref{eq:Reaction_rate_from_Floquet_exponents}
for obtaining rates directly from the geometric stability (and instability)
of the \ac{TS} trajectory.
{Calculating the reaction rates with the Floquet method is about two 
orders of magnitudes faster in computational time than the ensemble method. This
relies on the fact that ensembles require the calculation of millions
of trajectories, while only single trajectories
need to be considered in the Floquet method.}

\section{Conclusion and outlook}
\label{sec:Conclusion}
In this paper, we have applied \acp{NN} to describe the \acp{NHIM} of
periodically driven multidimensional systems with a
rank-1 saddle.
Use of these \acp{NN} allows for the efficient
propagation of stabilized
trajectories on the \ac{NHIM} with respect to the unstable degrees
of freedom of a rank-1 barrier even for longer integration times.
Without such stabilization, these trajectories would
depart exponentially fast from the \ac{NHIM} to either the reactant
or the product side due to the limited precision of numerical calculations, 
as seen in in \REF~\onlinecite{hern19a}.
It enables the analysis of the long-time
dynamics on the \ac{NHIM} using a stroboscopic \ac{PSOS}.
Therein, fixed points can be determined by two different methods,
viz.\ the centroid search and the friction search.
The application of the NN-enabled approaches to
a model system with two degrees of freedom
reveals the
existence of near-integrable tori surrounding 
a periodic \ac{TS}
trajectory given by a period one fixed point of the Poincar\'e map.

The rate constants of the system 
obtained from
the propagation of
large trajectory ensembles are in excellent agreement with the
rates obtained by a
Floquet analysis
associated with the periodic TS trajectory
extended and applied to
a periodic
\ac{TS} trajectory of the higher-dimensional system.
In \REF~\onlinecite{hern14f} 
this analysis was applied to the periodic
\ac{TS} trajectory of a \emph{one}-dimensional system, which coincides with
the \ac{NHIM}.
Here, we have generalized this method to any periodic trajectory
on the \ac{NHIM} of a \emph{multidimensional} rank-1 saddle.

While the model system shows regular dynamics on the \ac{NHIM},
it remains for future work to investigate whether a
transition from regular to chaotic dynamics can be observed, e.g.,
when changing the amplitude and frequency of the moving saddle.
It will also be challenging to study rate constants for trajectories
crossing the \ac{DS} not close to the periodic \ac{TS} trajectory but in other
regions of phase space.

Finally, the methods introduced in this paper should be applied to
other more realistic multidimensional systems, like the LiCN
$\leftrightarrow$ LiNC \cite{hern16c} or the ketene \cite{hern16d}
isomerization reactions, and further extended to systems with thermal
activation or friction.

\vspace{0pt}\begin{acknowledgments}

The German portion of this collaborative work was supported
by Deutsche Forschungsgemeinschaft (DFG) through Grant
No.~MA1639/14-1.
RH's contribution to this work was supported by the National Science
Foundation (NSF) through Grant No.~CHE-1700749.
MF is grateful for support from the Landesgraduiertenf\"orderung of
the Land Baden-W\"urttemberg.
This collaboration has also benefited from support by the European
Union's Horizon 2020 Research and Innovation Program under the Marie
Sklodowska-Curie Grant Agreement No.~734557.
The machine learning framework \emph{Tensorflow} has been used
to train the neural networks~\cite{tensorflow2015-whitepaper}.
\end{acknowledgments}

\bibliography{q17bib}

\end{document}